\documentstyle[12pt,aaspp4,flushrt]{article}

\makeatletter

\@ifundefined{chapter}{\def\thebibliography#1{\section*{References\@mkboth
  {REFERENCES}{REFERENCES}}\list
  {\relax}{\setlength{\labelsep}{0em}
        \setlength{\itemindent}{-\bibhang}
        \setlength{\itemsep}{0pt}
        \setlength{\parsep}{0pt}
        \setlength{\leftmargin}{\bibhang}}
    \def\newblock{\hskip .11em plus .33em minus .07em}
    \sloppy\clubpenalty4000\widowpenalty4000
    \sfcode`\.=1000\relax}}%
{\def\thebibliography#1{\chapter*{Bibliography\@mkboth
  {BIBLIOGRAPHY}{BIBLIOGRAPHY}}\list
  {\relax}{\setlength{\labelsep}{0em}
        \setlength{\itemindent}{-\bibhang}
        \setlength{\itemsep}{0pt}
        \setlength{\parsep}{0pt}
        \setlength{\leftmargin}{\bibhang}}
    \def\newblock{\hskip .11em plus .33em minus .07em}
    \sloppy\clubpenalty4000\widowpenalty4000
    \sfcode`\.=1000\relax}}

\newlength{\bibhang}
\let\@internalcite\cite
\def\cite{\let\@citeleft(\let\@citeright)%
    \@ifstar{\citeyear}{\citefull}}
\def\citenp{\let\@citeleft\relax\let\@citeright\relax
    \@ifstar{\citeyear}{\citefull}}
\def\citefull{\def\astroncite##1##2{##1~##2}\@internalcite}
\def\citeyear{\def\astroncite##1##2{##2}\@internalcite}

\def\@citex[#1]#2{\if@filesw\immediate\write\@auxout{\string\citation{#2}}\fi
  \def\@citea{}\@cite{\@for\@citeb:=#2\do
    {\@citea\def\@citea{; }\@ifundefined
       {b@\@citeb}{{\bf ?}\@warning
       {Citation `\@citeb' on page \thepage \space undefined}}%
{\csname b@\@citeb\endcsname}}}{#1}}

\def\@cite#1#2{\@citeleft#1\if@tempswa , #2\fi\@citeright}
\def\@biblabel#1{}

\makeatother

\newcommand{\PSbox}[3]{\mbox{\rule{0in}{#3}\includegraphics{#1}\hspace{#2}}}
\newcommand{\FigNum}[1]{\unitlength 1pt \begin{picture}(55,10)(-400,35) 
                        \put(0,0){Figure #1}
                        \end{picture}}

\def\persec{{s$^{-1}$}}
\def\etal{{\rm et~al.\/}}

\newcommand\approxlt{\mbox{$^{<}\hspace{-0.24cm}_{\sim}$}}

\begin{document}

\title{The $b$ Distribution and the Velocity Structure of Absorption
Peaks in the Lyman-Alpha Forest}
\author{Lam Hui\altaffilmark{1} and Robert E.
Rutledge\altaffilmark{2}}
\altaffiltext{1}{NASA/Fermilab Astrophysics Center, Fermi
National Accelerator Laboratory, Batavia, IL 60510; e-mail: \it lhui@fnal.gov}
\altaffiltext{2}{Max-Planck-Institut f\"{u}r Extraterrestrische
Physik, Postfach 1603, D-85740 Garching, Germany; e-mail: \it
rutledge@rosat.mpe-garching.mpg.de}

\begin{abstract} 

{A theory is developed which relates the observed $b$-parameter of a
Ly$\alpha$ absorption line to the velocity-curvature of the corresponding peak
in the optical depth fluctuation. Its relation to the traditional
interpretation of $b$ as the thermal broadening width is discussed.
It is demonstrated that, independent
of the details of the cosmological model, the differential $b$
distribution has a high $b$ asymptote of $dN/db \propto b^{-m}$, where
$m \geq 5$, when we make the reasonable assumption that low-curvature
fluctuations are statistically favored over high-curvature
ones. There in general always exist lines much broader than the
thermal width. We develop a linear  
perturbative analysis of the optical depth fluctuation, which yields
a single-parameter prediction for the full $b$ distribution. In
addition to exhibiting the high velocity tail, it 
qualitatively explains the observed  
sharp low $b$ cut-off -- a simple reflection of the
fact that high-curvature fluctuations are relatively rare.  
While the existence of the high $b$ asymptote, which is
independent of the validity of the linear expansion, is
consistent with the observed $b$ distribution, 
a detailed comparison of the linear prediction with six observational
datasets indicates that higher order corrections are not negligible.
The perturbative analysis nonetheless offers valuable insights into
the dependence of the $b$ distribution on cosmological
parameters such as $\Omega$ and the power spectrum. A key parameter is the
effective smoothing scale of the optical depth fluctuation, which is
in turn determined by three scales: the thermal broadening
width, the baryon smoothing scale (approximately the Jeans scale) and
the observation/simulation resolution. The first two are determined by
reionization history, but are comparable in general, while the third 
varies by about an order of magnitude in current hydrodynamic
simulations. Studies with non-resolution-dominated $b$ distributions
can be used to probe the reionization history of the universe.
}
\end{abstract}


\section{Introduction}
It was recently pointed out by Rutledge \cite*{rutledge97a} that
structures which arise naturally in hierarchical clustering models
imply a high-velocity tail for the $b$-parameter distribution of the
Ly$\alpha$ forest. The analysis was done using idealized filaments and
pancakes. We show in this paper that the same qualitative (but
different quantitative) behavior can be understood in the broader
context of the statistics of peaks in the optical depth fluctuation.

Both the high $b$ tail as well as the sharp low $b$ cut-off in the observed 
$b$-parameter distribution have been noted in
the literature (e.g. \citenp{press93}; for recent results, see
\citenp{hu95}, \citenp{lu96} and \citenp{kirkman97}). Although there
are subtle effects on the detection of narrow and broad lines due to
finite $S/N$ and continuum fitting (\citenp{rauch93}), the above
features seem to be robust (see the above ref.). 
Recent hydrodynamic simulations and semi-analytical calculations
reproduce the same features 
(\citenp{cen94}, \citenp{zhang95}, \citenp{hernquist96},
\citenp{jordi96}, \citenp{zhang96}, \citenp{bi97} and
\citenp{dave97}; but see also \citenp{haehnelt97}), although they do 
differ in details, a point to which we will return.

In \S \ref{high}, we propose relating the measured $b$-parameter of a
Ly$\alpha$ absorption line to the curvature around a peak in the
optical depth fluctuation. Under very general conditions, a high $b$
tail of $dN/db \propto b^{-m}$ is predicted for the differential $b$
distribution. The viewpoint adopted here is that the low column density
Ly$\alpha$ forest
($N_{\rm HI} \, \approxlt \, 10^{14} \, {\rm \, cm^{-2}}$)
is part of a fluctuating intergalactic medium, as predicted by
structure formation models (\citenp{bbc92,rm95}).  In \S
\ref{thermal}, we discuss the physical meaning of the {\it measured}
$b$ in the context of such models and its relationship with the
traditional interpretation of $b$ as the thermal (plus turbulence)
broadening width.  We illustrate these ideas by giving a concrete
example in \S \ref{linear}.  A linear perturbative expansion of the
optical depth fluctuation is developed, which yields a simple
prediction for the full $b$ distribution: in addition to showing the
high $b$ tail, it also implies a sharp low $b$ cut-off similar to that
found in the observed distribution. The single-parameter
linear prediction, as opposed to the commonly used three-parameter
truncated Gaussian, is compared with the $b$ distributions of
six datasets. Four of them can be satisfactorily described
by the model while the other two cannot.
We discuss the possible causes, and argue the main reason 
is that higher order corrections are
non-negligible. Nonetheless, the linear
analysis provides us useful intuition on how the velocity structure of
absorption lines reveals or depends on the cosmology, thermal history
and resolution of observations/simulations.  Finally we conclude in \S
\ref{discuss}.

\section{The High $b$ Tail}
\label{high}
Let us expand the optical depth $\tau$ as a function of
velocity $u$ around an absorption line:
\begin{equation}
\tau (u) = {\rm exp}\left[{{\rm ln}\,\tau(u)} \right]= \tau(u_{\rm
max}) \, {\rm exp}\left[{{1\over 
2}[{\rm ln}\, \tau]'' (u-u_{\rm max})^2}\right] \, ,
\label{tauexpand}
\end{equation}
where $u_{\rm max}$ is the velocity coordinate of the line center, and
the prime denotes differentiation with respect to $u$ and the second
derivative is evaluated
at $u_{\rm max}$. The
first derivative vanishes because $\tau$ is at a local
extremum. The reader will
recognize the above expansion as none other
than the Voigt profile: $\propto {\rm exp} [-(u-u_{\rm max})^2/b^2]$. 
The simple thermal profile suffices for our purpose, since we will focus
on the low column density Ly$\alpha$ forest.
A similar expansion, applied to the density field in comoving space (as
opposed to the optical depth field in velocity space here),
has been used to study the column density distribution
(\citenp{gh96,hui96b}). 

In other words, $b = \sqrt{-2/[{\rm ln}\, \tau]''}$. No
assumption has been made about whether the given absorption line is
thermally broadened or not. We will refer to it as the $b$-parameter,
or more simply $b$, instead of the commonly used Doppler parameter. As we
will show in \S \ref{thermal}, the measured $b$ agrees with the thermal
$b$ ($\propto \sqrt T$) only if the neutral hydrogen density is
sharply peaked in velocity space.  
Let us denote ${\rm ln}\, \tau$ by $\tau_L$. The number density of peaks
in $\tau_L$ (or minima in transmitted flux) with a given $b$ or
$\tau_L''$ is:
\begin{eqnarray}
{d N_u \over d b} = {4\over b^3} {d N_u \over {d\tau_L''}} = {4\over
b^3} |\tau_L''| P(\tau_L'=0,\tau_L'') \, \, \, \, , \, \, \, \, \tau_L'' = -{2
\over b^2}
\label{dnudb}
\end{eqnarray}
where $N_u$ is the number of peaks in $\tau_L$ per unit $u$, and
$P(\tau_L',\tau_L'') d\tau_L' d\tau_L''$ is the probability at any
given point along the spectrum that $\tau_L'$ and 
$\tau_L''$ take the values in the prescribed ranges. The equality
relating $d N_u/ d\tau_L''$ and $P$ above
follows from this argument: the number density of peaks
for a particular realization of the spectrum is a sum of Dirac delta
functions $\sum_i \delta_D (u-u_i)$ where $i$ denotes all the places
where $\tau_L$ is at a local maximum with a given $\tau_L''$; close to
a local maximum, $\tau_L'(u) \sim (u-u_i) \tau_L''(u_i)$; changing the
independent variable of the delta function from $u$ to $\tau_L'$ and
performing an ensemble average yields the above result (see \citenp{bbks86}).

It can be seen that as long as $P(\tau_L'=0,\tau_L'')$ approaches a
finite non-zero limit as $|\tau_L''|$ approaches $0$ (large $b$),
eq. (\ref{dnudb}) implies $dN_u /db \propto b^{-5}$ for large $b$. 

At any point along the line of sight where the first derivative vanishes,
a large curvature (or absolute value of the second derivative)
of random fields of cosmological interest, like the optical depth
or density fluctuation, should correspond to 
a rare event (being correlated with high maxima, which are themselves
rare) while a small value is common-place. 
The $b^{-5}$ asymptote is then a non-trivial consequence
of the expectation that small $|\tau_L''|$'s are not rare (in the sense that
$P[\tau_L'=0,\tau_L''=0]$ is a finite non-zero number).

The above argument is not completely general, however, 
because the condition that $P[\tau_L'=0,\tau_L''=0]$ is finite and
non-vanishing is not the only possible quantitative manifestation
of the expectation that $|\tau_L''|$'s are not rare.
More generally, we can state the following condition:
that $P_f [\tau_L'=0,f(\tau_L'')]$ approaches a finite non-zero value
as $f$ approaches $f(\tau_L''=0)$, where $f$ is a smooth analytic function of
$\tau_L''$ around $\tau_L''=0$ and $P_f (\tau_L'=0,f) d\tau_L' df$ is now the
probability that  
$\tau_L'$ and $f$ take the prescribed values.  We allow the
possibility that the old probability density $P(\tau_L'=0,\tau_L'')$
vanishes at $\tau_L'' = 0$, but assume it is non-singular at that point. 
One can 
replace $P$ in eq. (\ref{dnudb}) by $P_f df/d\tau_L''$ and Taylor expand $f$
around $\tau_L'' = 0$ using 
$f = \sum_{i=0}^{\infty} f_i \tau_L''^i$. 
Eq. (\ref{dnudb}) then implies that $dN_u /db \propto b^{-m}$ for sufficiently
large $b$, with $m \geq 5$.
For instance, suppose $f_i = 0$ for all $i$ less than some number $n$
(excluding $f_{i=0}$, which is irrelevant for $P_f df/d\tau_L''$) and
$f_i \neq 0$ for $i = n$ (with no additional condition imposed on all other $f_i$'s),
then $dN_u /db \propto b^{-m}$ in the large $b$ limit, with $m = 5+2(n-1)$. 
\footnote{The analyticity condition on $f$ is an
important one which allows us to Taylor expand. Consider a counter-example where $f =
{\tau_L''}^{k}$ with $k < 1$, then $df/d\tau_L''$ diverges at $\tau_L'' =
0$, and our condition that $P_f [\tau_L'=0,f(\tau_L''=0)]$ is finite
would imply the old 
probability density $P(\tau_L'=0,\tau_L'')$ diverges at $\tau_L'' =
0$. We take it to be a plausible assumption that
$P(\tau_L'=0,\tau_L'')$ is non-singular.  }

The $b^{-5}$ tail mentioned earlier corresponds to the case where $f_{1}$ is
non-vanishing. We will show in \S \ref{linear} that this is the case
for Gaussian random optical depth fluctuations.

How large does $b$ have to be for the $b^{-m}$ asymptote to take over? 
One can define the following velocity scale:
\begin{equation}
b_{\rm high} = \sqrt{P_f^{-1} {\partial \over \partial \tau_L''} \left[{P \over
{n f_n \tau_L''^{n-1}} }\right]}\, ,
\label{bhigh}
\end{equation}
where all terms are evaluated at vanishing $\tau_L'$ and $\tau_L''$, and
$f_{i=n}$ is the first non-vanishing $f_i$ defined before.
The quantity $b_{\rm high}$ gives us an estimate of how large $b$ has to
be for the onset of the $b^{-m}$ asymptote.
In other words, it tells us how low $\tau_L''$ has to be
for $P/(n f_n \tau_L''^{n-1})$ to be well-approximated by a simple
non-vanishing constant.
Note that in cases where the partial derivative with respect to $\tau_L''$ in
eq. (\ref{bhigh}) vanishes,  
for instance in the linear theory of \S \ref{linear}, 
analogous quantities can be defined using higher derivatives.

How does the above theoretical prediction fare with observations? 

From a sample of $790$ Ly$\alpha$ lines with $12.5 \le \log(N_{\rm
HI}) \le 14$ taken from the published line-lists of Hu et
al. \cite*{hu95}, we compute the cumulative $b$ distribution and
compare it with a single power-law $dN/db \propto b^{-m}$ using only
lines above a value 
$b_{\rm cut}$ using a Kolmogorov-Smirnov test \cite{press}, which produces the
probability (${\rm p_{\rm KS}}$) that a realization of the theoretical
distribution would produce a dataset which is more disparate than the observed
distribution is from the theoretical one.  We note that the four lines of
sights in the sample all have 
similar redshift ranges (see Table~\ref{tab:dataset}).  For lines with $b>30$
km \persec (368 
lines), the ${\rm p_{\rm KS}}>0.005$ range (at 0.01 increments) for $m$ is
[-4.12,-3.04];  for $b>40$ km \persec (180 lines), it is
[-5.03,-3.11]; and for $b>50$ km \persec (94 lines), it is
[-6.10,-2.89]. We note that restricting the $b$ range to higher
values, one eventually can obtain a sample which is consistent with
any power-law slope.  Thus while the observed distribution is
consistent with the $-m$ ($m \geq 5$) power-law for sufficiently high
$b$'s, it does not require one.  Within the context of our theory,
$\sim 40 \, {\rm km \, s^{-1}}$ can be regarded as a lower-limit to
$b_{\rm high}$ defined in eq. (\ref{bhigh}).

Before we move on to the next section, an important caveat on the peak
picture of an absorption line: in hierarchical clustering models, a
given peak in the optical depth does
not, in general, have an exact Voigt-profile shape. Standard profile fitting
routines might fit the peak with a profile as described in eq.
(\ref{tauexpand}), together with a few smaller profiles to fill in the
``wings'' of the peak. If 
this occurs frequently enough, a prediction based on eq. (\ref{dnudb})
might fail to match the observed $b$ distribution. However, the
success of the peak picture in another context, the column density
distribution, lends support to it.  We will take it to be our working
hypothesis, which can be checked through detailed comparisons with
simulations. It should be noted also that higher order terms in the Taylor
series expansion in eq. (\ref{tauexpand}) contain information about
departure from the Voigt-profile shape, and their statistics
could in principle be computed. We leave it for future work.

\section{Thermally Broadened -- or Not?}
\label{thermal}

The high velocity tail discussed in the last section implies that
there always exist very broad absorption lines. An obvious question:
should the high $b$ value be taken to indicate high
temperature, as is traditionally assumed? 

The answer is: not necessarily. In general, there inevitably exist
absorption lines much broader than the thermal broadening width. 
Let us try to understand its physical
origin. The optical depth $\tau$ at a given velocity $u_0$ is given by
(see \citenp{hui96b}):
\begin{equation}
\tau(u_0) = \sum \int {n_{\rm HI} \over {1+\bar z}} \left\vert {du\over
dx}\right\vert^{-1} 
\sigma_\alpha du \,\,  , \,\, \sigma_\alpha = \sigma_{\alpha 0}
{c\over {{b_T}\sqrt{\pi}}} 
\, {\rm exp} [{-{(u-u_0)^2/{b_T}^2}}] \, ,
\label{tau}
\end{equation}
where $n_{\rm HI}$ is the proper number density of neutral hydrogen, $\bar z$
is the mean redshift of interest, $x$ is the comoving spatial
coordinate and the integration is done over the velocity $u$ along the
line of sight. The Jacobian $|du/dx|$ multiplying the proper density
$n_{\rm HI}$  gives us the neutral hydrogen density in velocity-space, and the summation
is over multiple streams of $x$'s at a given $u$.

The thermal profile is given in the second equality, with
$\sigma_{\alpha 0}$ being the Lyman-alpha cross section constant
(\citenp{rybicki97}). 
The width of the profile is $b_T = \sqrt{2 k_B T / m_p}$
where $T$ is the temperature of the 
gas, $k_B$ is the Boltzmann constant and $m_p$ is the mass of a
proton. Turbulence broadening can in principle be included simply
by defining an effective temperature ($T = T_{\rm turb.} + T_{\rm
thermal}$), and we will use the term thermal broadening to refer to both.

It can be seen from eq. (\ref{tau}) that if the HI number density in
redshift space $n_{\rm HI} |du/dx|$ is
sharply peaked (e.g. a Dirac delta function), $\tau$ has the shape
of the thermal profile around the peak, and the measured $b$ as
proposed in eq. (\ref{tauexpand}) would coincide with the width $b_T$.
However, in the opposite limit in which $n_{\rm HI} |du/dx|$ 
is varying slowly around a peak, $n_{\rm HI} |du/dx|$ can be
taken out of the integral, and the result is that $\tau$ does not
have the shape of the thermal profile in general, and its width is determined
by the scale of variation of $n_{\rm HI} |du/dx|$, which is larger
than $b_T$ in this case. The {\it measured} $b$ defined in eq.
(\ref{tauexpand}) then
reflects the velocity structure of the peak in HI number density,
rather than the temperature of the gas. 
The existence of absorption lines with high $b$ ($ > b_T$) is
unavoidable as long as 
structure formation models allow fluctuations on large scales, in other words,
low-curvature fluctuations. \footnote{Press \& Rybicki \cite*{press93} arrived
at the conclusion that the thermal interpretation of the observed high $b$
values is inconsistent with the hypothesis of heating and photoionization
equilibrium by a UV background using a different argument: essentially baryon
counting.}

The arguments above set $b_T$ as the lower limit to the observed
width of an absorption line, but there always exist lines which are
much broader.
Moreover, if the HI number density in
redshift space is intrinsically smooth on scales larger than $b_T$,
{\it most} lines would in fact be wider than the thermal broadening width.
Possible sources of such smoothing are the Jeans-smoothing and the effective
numerical/observational resolution, to which we will return.

In the next section, we develop a linear perturbative analysis of the
optical depth fluctuation (small fluctuation limit), which exhibits
all these possibilities, and allows us to predict the full $b$
distribution, from narrow to broad lines. It will be demonstrated that
the temperature of the intergalactic medium influences the overall $b$
distribution not so much by giving a thermal width to every absorption
line, but by determining the amount of smoothing of the optical depth
field, which affects the relative probability of high-curvature versus
low-curvature fluctuations. We emphasize, however, that the key 
result in \S \ref{high}, namely the existence of the $b^{-m}$ asymptote, is
independent of the validity of the linear calculation.

\section{A Linear Analysis of the Optical Depth Fluctuation}
\label{linear}

Assuming that $\tau(u) = \bar\tau [1 + \delta_\tau (u)]$ where
$\bar\tau$ is the mean optical depth and $\delta_\tau
\ll 1$, the quantities
$\tau_L'$ and $\tau_L''$ in the expression for 
$dN_u/db$ in eq. (\ref{dnudb}) can be replaced by $\delta_\tau'$ and
$\delta_\tau''$ respectively.

As we will show below, for cosmological models with Gaussian random
initial conditions, the linear fluctuation $\delta_\tau(u)$ is itself Gaussian
random, 
and so the probability distribution $P(\delta_\tau'=0,
\delta_\tau'')$ is given by the product of two Gaussians,
with $\delta_\tau'$ set to 0: 
$(2\pi
\sqrt{\langle\delta_\tau'^2\rangle
\langle\delta_\tau''^2\rangle})^{-1}$ ${\rm exp}[-\delta_\tau''^2/(2 
\langle\delta_\tau''^2\rangle)]$, where $\langle\rangle$ denotes
ensemble averaging. The normalized
$b$ distribution then follows from eq. (\ref{dnudb}):
\begin{equation}
{d N \over db} = {4 b_\sigma^4 \over b^5} {\rm exp}\left[{-{b_\sigma^4
\over b^4}}\right]
\, \, , \, \, b_\sigma^4
\equiv {2 \over {\langle{{\delta_\tau}''}^2\rangle}} \, ,
\label{dNdb}
\end{equation}
where the normalization is chosen such that $\int_0^{\infty} (dN/db) db
= 1$.\footnote{The non-normalized distribution, or the
total number of absorption lines, is interesting in its own right. It turns
out that to model it properly, at least one important selection effect has to 
be taken into account, namely typically only lines above a certain
column density are included. The above
formalism can be modified to accommodate it by including a threshold
to $\delta_\tau$, which is correlated with $\delta_\tau''$. 
Such a modification would in fact alter the shape of the normalized
$b$ distribution as well. We leave complications due to this
selection effect as well as others for a future paper.}

The $b^{-5}$ high $b$ asymptote is clearly exhibited. As is shown in
\S \ref{high}, this is a result of the fact that $P(\delta_\tau'=0,
\delta_\tau'')$, which is proportional to ${\rm exp}[-b_\sigma^4/b^4]$
here, approaches a finite constant in the large $b$ limit. 

This linear Gaussian model also predicts the presence of a very sharp
cut-off at low-$b$'s; such a cutoff has been reported from
observations \cite{hu95,lu96,kirkman97,kim97}.  This distribution,
with $b_\sigma=26.3$ km \persec , is shown super-imposed upon the
observed distribution from one line of sight (QSO 0014+813) in
Fig.~\ref{fig:dist}. Note how a single parameter $b_\sigma$ controls
both where the low $b$ cut-off and the high $b$ asymptote take over. It is
determined by the rms fluctuation amplitude of $\delta_\tau''$.  Let
us derive its dependence on cosmological parameters.

Returning to eq. (\ref{tau}), let us first note that in the linear regime, one
can ignore multiple streaming and therefore the summation.
Ionization equilibrium implies that the $n_{\rm HI}$ is determined by
the local baryon overdensity, let us 
call it $\delta$, and the local temperature through $n_{\rm HI} \propto
[1+\delta]^2 T^{-0.7}$. 
The temperature $T$ is typically related to $\delta$
by $T = T_0 (1+\delta)^{\gamma-1}$, where $T_0$ is the mean temperature
at $\delta=0$ and $\gamma$ is determined by reionization history
(\citenp{hui96a}). The velocity $u$ is related to the spatial
coordinate $x$ by $u = H (x - \bar x)/ (1+\bar z) + v_{\rm pec}$
where $\bar x$ is the mean position of interest and $H$ is the
Hubble constant at redshift $\bar z$, and $v_{\rm pec}$ is the
peculiar velocity along the line of sight. 

Collectively, the above relations imply that the optical depth
fluctuation $\delta_\tau$ is simply determined by two random fields: $\delta$
and $v_{\rm pec}$. The reader is referred to Hui et al. \cite*{hui96b}
for details. Keeping only terms to first order in $\delta$ and $v_{\rm
pec}$, one obtains from eq. (\ref{tau}):
\begin{eqnarray}
&&\delta_\tau (u_0)  = \int \Biggl[
[2-0.7(\gamma-1)]\delta - {\partial v_{\rm 
pec}\over \partial u} + (\gamma-1) {b_{T_0}^2 \over 4} {\partial^2 \delta \over
\partial u^2}
\Biggr] W(u-u_0) du \, \, , \nonumber \\ &&W(u-u_0) \equiv {1\over 
{b_{\rm eff}} \sqrt{\pi}} \, {\rm exp}
[-{(u-u_0)^2/b_{\rm 
eff}^2}] \, ,
\label{deltatau}
\end{eqnarray} 
where $\delta_\tau = (\tau - \bar\tau)/\tau$, and $b_{T_0} = \sqrt{2
k_B T_0 / m_p}$ is the thermal broadening width at temperature $T_0$,
and $W$ is simply a Gaussian smoothing window. The quantity $b_{\rm
eff}$ should be set to $b_{T_0}$ strictly speaking, if one were to
derive the above from eq. (\ref{tau}). However, we allow them to be
different for the following reasons: 1. the observed spectrum is often
convoluted with a Gaussian resolution window (strictly speaking the
convolution is operated on ${\rm exp}(-\tau)$, but in linear theory
the same convolution is applied to $\delta_\tau$); one can also model
the resolution of the simulated spectrum similarly; 2. to relate the
above quantity directly to cosmology, it is convenient to replace the
baryon overdensity $\delta$ by the dark matter overdensity (assuming
the universe is dark matter dominated), which we will denote by
$\delta$ from now on, but a smoothing kernel has to be applied to the
latter to take into account the smoothing of small scale baryon fluctuations
due to finite gas pressure, which
can also be approximated by a Gaussian (\citenp{hui96a}) (similarly
for the $v_{\rm pec}$ field, which is related to $\delta$ by $v_{\rm
pec} = - \partial \nabla^{-2} 
\dot\delta/ \partial x$ in linear theory, where the dot denotes
differentiation with respect to conformal time).  The combination of
the three different Gaussian kernels, due to thermal broadening,
resolution of observation/simulation and smoothing by finite gas
pressure, is itself a Gaussian with width $b_{\rm eff}$ given by:
\begin{eqnarray}
&&b^2_{\rm eff} = b_{T_0}^2 + b_{\rm res}^2 + b_{\rm J}^2 \quad ,
\quad \quad {\rm
where} \nonumber \\
&&b_{T_0} = 13 \, {\rm km \, s^{-1}} \left[T_0 \over {10^4 {\rm
K}}\right]^{1\over 2} \, 
, \, 
b_{\rm res} = {{\rm FWHM}\over {2 \sqrt{{\rm ln\, 2}}}} \, , \,
b_{\rm J} = 24 \, 
{\rm km \, s^{-1}} f_{\rm J} \left[\gamma \over 1.5\right]^{1\over 2}
\left[T_0 \over 
{10^4 {\rm 
K}}\right]^{1\over 2} \, ,
\label{beff}
\end{eqnarray}
where $b_{T_0}$ is the thermal broadening width given before, 
$b_{\rm res}$ is the resolution width which is related to the commonly
quoted resolution FWHM by the factor given above, and $b_{\rm J}$ is
the baryon smoothing scale. 
If the factor $f_{\rm J}$ were set to $1$, $b_{\rm J}$ is exactly the
Jeans scale in 
the appropriate velocity units ($b_{\rm J} = 2 f_{\rm J} H k_{\rm
J}^{-1}/(1+\bar z)$ in the convention of \citenp{gh97}). 
 However, as shown by Gnedin \& Hui 
\cite*{gh97},  
the correct linear smoothing scale is in general smaller than, but of
the order of, the Jeans scale, if one waits for sufficiently long after
reionization. 
For example, if reionization occurs at $z=7$, 
$f_{\rm J} \sim
0.5$ at $z=3$. It is interesting to note that both $b_{T_0}$ and $b_{\rm J}$ 
scale as the square root of the temperature. 

To establish the Gaussianity of $\delta_\tau (u)$ in eq.
(\ref{deltatau}), assume that both fields
$\delta $ and $v_{\rm pec} $ are Gaussian random in the
three-dimensional comoving space
${\bf x}$, which is expected for a large class of 
inflationary cosmological 
models. One dimensional projection ${\bf x}$ to $x$ preserves 
Gaussianity, and so does the comoving space to redshift-space
mapping, to
the lowest order (i.e. one can equate a change of variable
from $u$ to $x$ or vice 
versa with the simple linear transformation $u
= H (x-\bar x)/(1+\bar z)$, if
one is only keeping first order terms as in eq. [\ref{deltatau}]).
These two facts mean that 
both $\delta$ and $v_{\rm pec}$ are Gaussian random in $u$ space. 
Finally, all operations on $\delta (u)$ and $v_{\rm pec} (u)$ in eq.
(\ref{deltatau}), differentiation, addition and the Gaussian
convolution, are linear which means $\delta_\tau (u)$ itself is
Gaussian random.

Therefore, simply put, $\delta_\tau$ is a Gaussian random field which
is equal to $[2-0.7(\gamma-1)]\delta - v' + (\gamma-1) {b_{T_0}^2
\delta'' /4}$, smoothed on the scale of $b_{\rm eff}$ given by eq.
(\ref{beff}). Similarly $\delta_\tau''$ is itself Gaussian random and
its statistical properties are completely specified by its two-point
function.

It is straightforward to compute $\langle \delta_\tau''^2 \rangle$, 
which gives $b_\sigma$ in eq. (\ref{dNdb}):
\begin{equation}
\langle \delta_\tau''^2 \rangle = {D_+^2}
\left[(1+\bar z) \over H\right]^{4}
\left[ ({1\over 5}\alpha^2 + {2\over 7}\alpha f_\Omega + {1\over 9}f_\Omega^2)
\sigma_2^2 - ({2\over 7} \alpha\eta + {2\over 9}f_\Omega\eta) \sigma_3^2 +
{1\over 9} \eta^2 \sigma_4^2 \right] \, ,
\label{rmsdt}
\end{equation}
where
\begin{equation}
\alpha \equiv 2-0.7(\gamma-1) \, \, , \, \, f_\Omega \equiv {d \, {\rm ln} \, D_+
\over {d \, {\rm ln} \, a}} \sim \Omega_m^{0.6}\, \, , \, \, \eta \equiv
b_{T_0}^2 {\gamma-1 \over 4} \left[1+\bar z \over H\right]^2 \, ,
\label{def}
\end{equation}
and 
\begin{equation}
\sigma_i^2 \equiv \int_0^{\infty} 4\pi k^{2i+2} P(k) \, {\rm
exp}\left[-{k^2\over k_{\rm 
eff}^2}\right] dk \, \, , \, \, k_{\rm eff} \equiv {\sqrt{2} H \over b_{\rm
eff} (1+\bar z)}  \, .
\label{sigmai}
\end{equation}
A few symbols require explanation: $D_+$ is the linear growth factor,
$a$ is the Hubble scale factor, $P(k)$ (to be distinguished from
$P[\tau_L',\tau_L'']$ used earlier) is the linear-extrapolated
power spectrum today as a 
function of the comoving wavenumber $k$, $\Omega_m$ is the matter
density, and redshift dependent 
quantities such as $H$ and $\Omega_m$ are to be evaluated at $\bar z$.

Given a cosmological model together with a reionization history,
$b_\sigma$ can be calculated from first principle, and eq.
(\ref{dNdb}) gives us the 
linear prediction for $dN/db$. Let
us see how the prediction fares with observations, and then sort out
the somewhat complicated (multi-) parameter dependence exhibited above.

\subsection{Comparison with the Observed Distribution}
\label{observe}

We compare the linear theory prediction for the differential $b$ distribution
in eq. (\ref{dNdb}) with datasets from six QSO lines of sight from three
different studies \cite{hu95,lu96,kirkman97}. 
The theoretical prediction is compared individually with each line of
sight, because in principle, the parameter $b_\sigma$ can evolve with
redshift.  In Table~\ref{tab:dataset}, we give the range of $b_\sigma$
with ${\rm p_{KS}}>$1\%, testing models with $b_\sigma$ in the range
10-30 km \persec, at 0.1 km \persec\ increments.  Ranges were found for
the four QSO datasets of Hu \etal \cite*{hu95}, with acceptable
$b_\sigma$ values in the range 23.3-28.8 km \persec. Of these four
datasets, three produced a range of $b_\sigma$ with ${\rm p}_{\rm KS}>10$\%.
Data from the QSO HS~1946+7658 line-of-sight \cite{kirkman97} and from
the Q0000-26 line-of-sight \cite{lu96} were not compatible with with
the theoretical distribution for the range of $b_\sigma$'s tested, with
maximum probabilities of ${\rm p_{KS}}=4\times10^{-3}$ and
$3\times10^{-3}$ at $b_\sigma=26.1$ and 23.7 km~\persec, respectively.
The differences found between the above studies could be partly a
result of different profile-fitting algorithms.

However, if we assume that $b_\sigma$ does not evolve significantly
over the redshift range covered by the Hu \etal \cite*{hu95} datasets
(they are at similar, but not exactly the same, redshifts),
grouping them together diminishes the agreement, and the
maximum ${\rm p_{KS}}$ reduces to $5
\times 10^{-4}$. We interpret the above findings mainly 
as an indication that higher
order corrections ignored in the linear analysis are not negligible,
as we will demonstrate in \S \ref{interprete} for a reasonable
cosmological model. 

Nonetheless, that a single value parametrization of the $b$
distribution can fit 
some of the datasets at all is interesting (as opposed to three
parameters in the case of the commonly used truncated Gaussian).
Let us also emphasize that the validity
of the peak picture in general, and the high $b$ asymptote in
particular, as discussed in \S \ref{high}, are independent of the
validity of the linear perturbative expansion.

\subsection{Physical Interpretation}
\label{interprete}

To quantify the departure from linearity, one can compute $\sqrt
{\langle \delta_\tau^2 \rangle}$ which is the rms fluctuation
amplitude of the optical depth $\delta_\tau$. For the CDM model with
$\sigma_8 = 0.7$, $\Omega_m = 1$, a present Hubble constant of $50 \,
{\rm km \, s^{-1} \, Mpc^{-1}}$, and assuming $b_{\rm eff} = 18 \,
{\rm km \, s^{-1}}$ (for resolution FWHM of about $8 \, {\rm km \,
s^{-1}}$, $T_0 = 10^4 K$, $\gamma = 1.5$ and $f_{\rm J}= 0.5$ in eq.
[\ref{beff}]), $\sqrt {\langle \delta_\tau^2 \rangle} = 3.6$ at
redshift $\bar z = 3$. The above parameters are known to give a column
density distribution that agrees with observations
(e.g. \citenp{hernquist96}, \citenp{zhang96} and \citenp{hui96b}).  It
is interesting to note that for the same parameters, the rms
fluctuation amplitude of the matter overdensity $\delta$ is $1.8$, a
factor of $2$ smaller: large scale coherent peculiar velocity flow
helps enhance the optical depth fluctuation.

Although the linear theory calculation fails, it may still help us
understand the broad-stroke parameter dependence of the $b$ distribution.
The key parameter in the calculation is $\langle \delta_\tau''^2
\rangle$, which gives an indication of how large the average $b$ is
expected to be, and how much spread one should expect in the $b$ distribution
(eq. [\ref{dNdb}]).

Let us dissect the parameter dependence of $\langle \delta_\tau''^2
\rangle$ in eq. (\ref{rmsdt}). Roughly speaking, $\sigma_i^2$ is of
the order of $\sigma_0^2 k_{\rm eff}^{2i}$ with $k_{\rm eff} \propto H/[b_{\rm
eff}(1+\bar z)]$ (see eq. [\ref{sigmai}]). Eq. (\ref{rmsdt}) and (\ref{dNdb}) then
imply
\begin{equation}
\sqrt {\langle \delta_\tau''^2 \rangle} \sim {D_+ \sigma_0 \over b_{\rm
eff}^2} \, \,  , \, \, b_{\sigma} \sim {b_{\rm eff}  \over \sqrt{D_+\sigma_0}}
\label{rmsdtAA}
\end{equation}
assuming that $\alpha$, $f_\Omega$ and $b_{T_0} / b_{\rm eff}$ are of the
order of unity. For the same CDM model discussed above, an exact calculation
yields $b_\sigma = 14 \,
{\rm km \, s^{-1}}$, which is rather low compared with the 
observed $b_\sigma$ in the range 23-29 km \persec. We will return to
this point below.

\begin{center}
{\it The Smoothing Scale}
\end{center}

Clearly, the one important parameter determining the relative number
of high $b$ to low $b$ absorption lines is the smoothing scale $b_{\rm
eff}$, which is the only relevant velocity scale in the problem.
The larger $b_{\rm eff}$ is, the broader the lines are in general. 
Note that increasing $b_{\rm eff}$ (decreasing
$k_{\rm eff}$) decreases $\sigma_0$ (eq. [\ref{sigmai}]), which
can only enhance the effect.

The effective smoothing scale $b_{\rm eff}$ has contributions from
three sources: thermal
broadening (including local turbulence) ($b_{T_0}$), observation/simulation
resolution ($b_{\rm res}$) and smoothing due to 
finite gas pressure ($\sim$ Jeans scale; $b_{\rm J}$) (eq. [\ref{beff}]). 
We can see from eq. (\ref{beff}) the 
baryon smoothing scale $b_{\rm J}$ is in fact of the same order as
$b_{T_0}$ for $f_{\rm J}
\sim 0.5$, unless turbulence is important (in which case the effective
$T_0$ going into $b_{T_0}$ should be higher than the thermal $T_0$
going into $b_{\rm J}$).
The precise value of $f_{\rm J}$ depends very much on
reionization history, the overall trend being the closer the epoch
of reionization is to the time of interest, the smaller $f_{\rm J}$ would
be (see \citenp{gh97}). At a redshift of $3$, if reionization occurs
before about redshift of $7$, $b_{T_0}$ is interestingly
close to $b_{\rm J}$. 

The resolution width $b_{\rm res}$, on the other hand, can vary by
large amounts. High quality Keck spectra have FWHM of $8 \, {\rm km \,
s^{-1}}$, which corresponds to $b_{\rm res} \, \sim 5 \, {\rm km \,
s^{-1}}$ i.e. the resolution effect is expected to be subdominant compared
to the other 
two sources of smoothing. However, current simulations have effective
resolution spanning about an order of magnitude: $b_{\rm res} \sim 3
\, {\rm km \, s^{-1}} - 30 \, {\rm km \, s^{-1}}$ (\citenp{cen94},
\citenp{hernquist96}, \citenp{zhang96}; see, in particular,
discussions in \citenp{bond97}), which means for some of them, $b_{\rm
res}$ could be the dominating influence on the average width of an
absorption line.  This might explain some of the differences seen in
their predicted $b$ distributions (\citenp{dave97} \&
\citenp{zhang96}), although differences in their Voigt-profile fitting
techniques and reionization histories certainly contribute. One
should also bear in mind that, in the presence of nonlinear
corrections, the proper smoothing scale would no longer be
exactly $b_{\rm eff}$ (even the Gaussian form of the smoothing would
be changed), but the overall trend that higher $b_{T_0}$ and/or
$b_{\rm res}$ and/or $b_{\rm J}$ imply broader lines is expected to
remain true.

In the context of the linear theory, one can see that if $b_{\rm
eff}$ is dominated by $b_{T_0}$, $b_\sigma$ would be of
the order of $b_{T_0}$ (eq. [\ref{rmsdtAA}], assuming $D_+ \sigma_0
\sim 1$). This means the bulk of the absorption lines would have
widths close to $b_{T_0}$, but the existence of much broader lines is
also inevitable 
(eq. [\ref{dNdb}]). 
Moreover, if $b_{\rm eff}$ is dominated by the baryon-smoothing scale or
resolution, most of the lines would end up having widths larger than 
$b_{T_0}$. 

The temperature of the
intergalactic medium should not be understood as influencing the
overall $b$ distribution by giving a thermal width to every individual
absorption line. Instead, 
it affects the $b$ distribution by determining the amount of smoothing
of the optical depth field (through $b_{T_0}$ and $b_{\rm J}$), which
controls the relative probability of high-curvature versus
low-curvature fluctuations (a higher $T_0$ favors the later, 
thereby shifting the overall distribution to higher $b$'s; see eq.
[\ref{dNdb}], 
[\ref{beff}] and [\ref{rmsdtAA}]).  

\begin{center}
{\it Normalization of the Linear Power Spectrum}
\end{center}

The other (dimensionless) parameter influencing the $b$ distribution
in the linear theory is $D_+ \sigma_0$ (eq.  [\ref{rmsdtAA}]), which
is simply the rms linear overdensity fluctuation at the redshift of
interest. The more linear power a given model has at the smoothing scale
$b_{\rm eff}$, the lower $b_\sigma$ is, which implies on the whole
narrower absorption lines. In other words, high-curvature fluctuations are more
probable for a model with larger linear power. 
Whether this trend remains true after
nonlinear corrections kick in is unclear without a detailed
calculation. From the discussion in \S \ref{high}, we know that $b =
\sqrt{2/|({\rm ln} \, \tau)''|}$, independent of the validity of the
linear expansion.  One can take the magnitude of $\langle ({\rm ln} \,
\tau'')^2 \rangle$ as an indication of how narrow the absorption lines
are. Expanding it beyond the lowest order, subject to the extremum
condition $\tau' = 0$, we have $\langle ({\rm ln} \, \tau'')^2 \rangle =
\langle \delta_\tau''^2 \rangle - 2\langle \delta_\tau''^2 \delta_\tau
\rangle$. The higher order corrections can in fact make $\langle ({\rm
ln} \, \tau'')^2 \rangle$ smaller than the linear theory
prediction. This is consistent with the fact that the linear
prediction for the CDM model discussed earlier is $b_\sigma = 14 \,
{\rm km \, s^{-1}}$ which is rather lower than the observed values
($\sim$ 23-29 km \persec ; one could of course argue the whole
perturbative expansion breaks down for the above model, but it might
work better for models with less small-scale power such as the mixed
dark matter models).  The overall effect of raising the normalization
of the linear power spectrum is therefore unclear. For the same
reason, it is difficult to make any qualitative statement regarding
the redshift evolution of the normalized $b$ distribution, because it is
unclear in what form the combination $D_+ \sigma_0$ would appear in the
expression for $dN/db$ if nonlinear corrections are included. On the other hand, one
other source of redshift evolution is the evolution of the effective smoothing
scale $b_{\rm eff}$ discussed earlier, which is mainly determined by 
the evolution of the equation of state of the intergalactic medium
i.e. reionization history.

\begin{center}
{\it $\Omega_m$}
\end{center}

Assuming fixed $b_{\rm eff}$, the parameter $\Omega_m$ enters into
the linear prediction for the $b$ distribution in two different
places: $D_+$ and $f_\Omega$ (eq. [\ref{rmsdt}] and [\ref{dNdb}]). 
Lowering $\Omega_m$ has two effects: it decreases the growth rate $\dot
D_+$ and makes $f_\Omega$, which quantifies how the
comoving-to-redshift-space mapping
enhances fluctuations, smaller. However, for the same reason as
discussed in the case of the power spectrum normalization,
the qualitative effect of nonlinear corrections is difficult to guess
without a detailed calculation. But the linear 
calculation suggests that the $\Omega_m$ dependence of the $b$ distribution 
is inevitably tangled with the dependence on power spectrum normalization. 
One should also keep in mind that at sufficiently high redshifts, 
$\Omega_m$ is close to $1$ whatever $\Omega_m$ is today. However, for
open universe models, with $\Omega_m$ of $0.3$ today, $\Omega_m = 0.6$
even at a redshift of $3$. 

\begin{center}
{\it J}
\end{center}

Lastly, one parameter that is curiously missing from the linear theory
calculation is the ionization background $J$. It is easy to see that,
quite generally in fact, since one is only interested in the shape of
$\tau$ around an absorption peak, the overall normalization of $J$
does not play a role in determining the $b$ distribution.
This is true, however, only in so far as the range of $J$ being
considered is small enough that it does not affect severely
the selection of absorption lines into one's sample (e.g. because of
finite signal to noise, etc).

\section{Discussion}
\label{discuss}

We develop in this paper a theory relating the $b$-parameter of a
Ly$\alpha$ absorption line to the curvature of a peak in the optical
depth fluctuation. It is shown in \S \ref{high}, under the condition that
low-curvature optical depth fluctuations are common, an asymptote of $dN/db
\propto b^{-m}$ is in general expected in the broad-line limit, with
$m \geq 5$. This
is independent of details of the cosmological model, or the smallness
of the fluctuation.
The observed $b$ distribution is consistent with the onset of the
$b^{-5}$ asymptote for $b$ larger than $b_{\rm high} \sim 40 \, {\rm
km \, s^{-1}}$ (eq. [\ref{bhigh}]).

A perturbative analysis of the optical depth fluctuation is developed,
that, in addition to illustrating the above asymptote, predicts a
sharp low $b$ cut-off. This is a reflection of the fact that
high-curvature fluctuations are strongly suppressed in a Gaussian
random field.  Of the six datasets that we compared with the linear
theoretical distribution, four were consistent with the theory (with
${\rm p}_{\rm KS}>$ 1\%) for a range of $b_\sigma$.  The other
two were consistent, at best, at the ${\rm p}_{\rm KS} \sim
0.3$\% level. The fact that a single value parametrization as
predicted by the linear analysis, as opposed to the
commonly used three-parameter truncated Gaussian, works for some of them is
interesting. However, we interpret the general disagreement as an
indication of non-negligible nonlinear corrections. 

The effective smoothing scale $b_{\rm eff}$, which arises from a
combination of thermal broadening (including turbulence), baryon smoothing
due to finite gas pressure and finite observation/simulation
resolution ($b_{T_0}$, $b_{\rm J}$, $b_{\rm res}$ in eq.
[\ref{beff}]), is the key parameter that determines 
the overall width 
of the absorption lines. The higher $b_{\rm eff}$ is, the broader the
absorption lines are in general. The baryon smoothing is shown to be at least as
important as thermal broadening for sufficiently early reionization,
unless turbulence is a significant source of broadening. If
reionization occurs very close to the redshift of interest, however,
$b_{T_0}$ could dominate, because it takes a while for $b_{\rm J}$ to
grow to an appreciable value after reionization (\citenp{gh97}). 

The importance of $b_{\rm res}$ varies a lot between different
observations/simulations. The resolution effect is probably
unimportant for high quality Keck data, while some of the current
hydrodynamic simulations might produce resolution-dominated $b$
distributions. The fact that Keck-quality data are not resolution-dominated
also means one can potentially use the observed $b$ distribution to
probe reionization history through its dependence on $b_{T_0}$ and
$b_{\rm J}$ (\citenp{haehnelt97}, see also \citenp{hui96a}). 
 
As we argue in \S \ref{thermal}, while the thermal broadening width does
provide a lower limit to the width of an absorption line, the
existence of
much broader lines is an inevitable consequence of the fact that
low-curvature fluctuations are statistically favored. 
The simple thermal interpretation of all the observed $b$ values is therefore
not viable, at least in the context of current structure formation models.
The temperature of the intergalactic medium affects the overall $b$
distribution not by giving a thermal width to every individual absorption line,
but by determining the amount of smoothing of the optical depth fluctuation.
Any attempt to understand the full $b$ distribution of the Ly$\alpha$ forest
and its dependence on reionization history must involve understanding the
statistics of the optical depth fluctuation, and its relation to
the underlying cosmological model. A first attempt has been made here
to explore how cosmological parameters such as $\Omega_m$ and the
power spectrum influence the $b$ distribution in the context of
linear theory. Further research taking into account nonlinear effects
is worth pursuing. The velocity structure of absorption peaks,
quantitied here by the $b$ value, may provide important constraints on
reionization history, as well as other cosmological parameters.

LH is supported by the DOE and by the NASA (NAGW-2381) at
Fermilab. RR is supported by a Max Planck Fellowship and is grateful for the
very kind hospitality of Prof. J. Tr\"umper during his visit to MPE. 
We thank Roman Scoccimarro for useful discussions and Martin Haehnelt
for helpful comments on our manuscript.


\begin{figure}[h]
\caption{ 
\label{fig:dist}  
 Comparison of the linear theory prediction for the differential
$b$-distribution, using $b_\sigma=26.3$ km sec$^{-1}$ (eq.
[\protect{\ref{dNdb}}] ) with the observed distribution of 194 Ly$\alpha$
lines with $10^{12.5} \le 
N_{\rm HI} \le 10^{14} \, {\rm cm^{-2}}$, from Q0014+813 (Hu \etal 1995; see
Table~\protect{\ref{tab:dataset}}) .  The theoretical distribution
qualitatively and quantitatively matches the data, with a cut-off at
low $b$ values, a quasi-Gaussian peak, and a high-$b$ tail.
Statistically, the fit is acceptable (${\rm prob}_{\rm KS}= 0.16$). }
\end{figure}

\begin{figure}
\PSbox{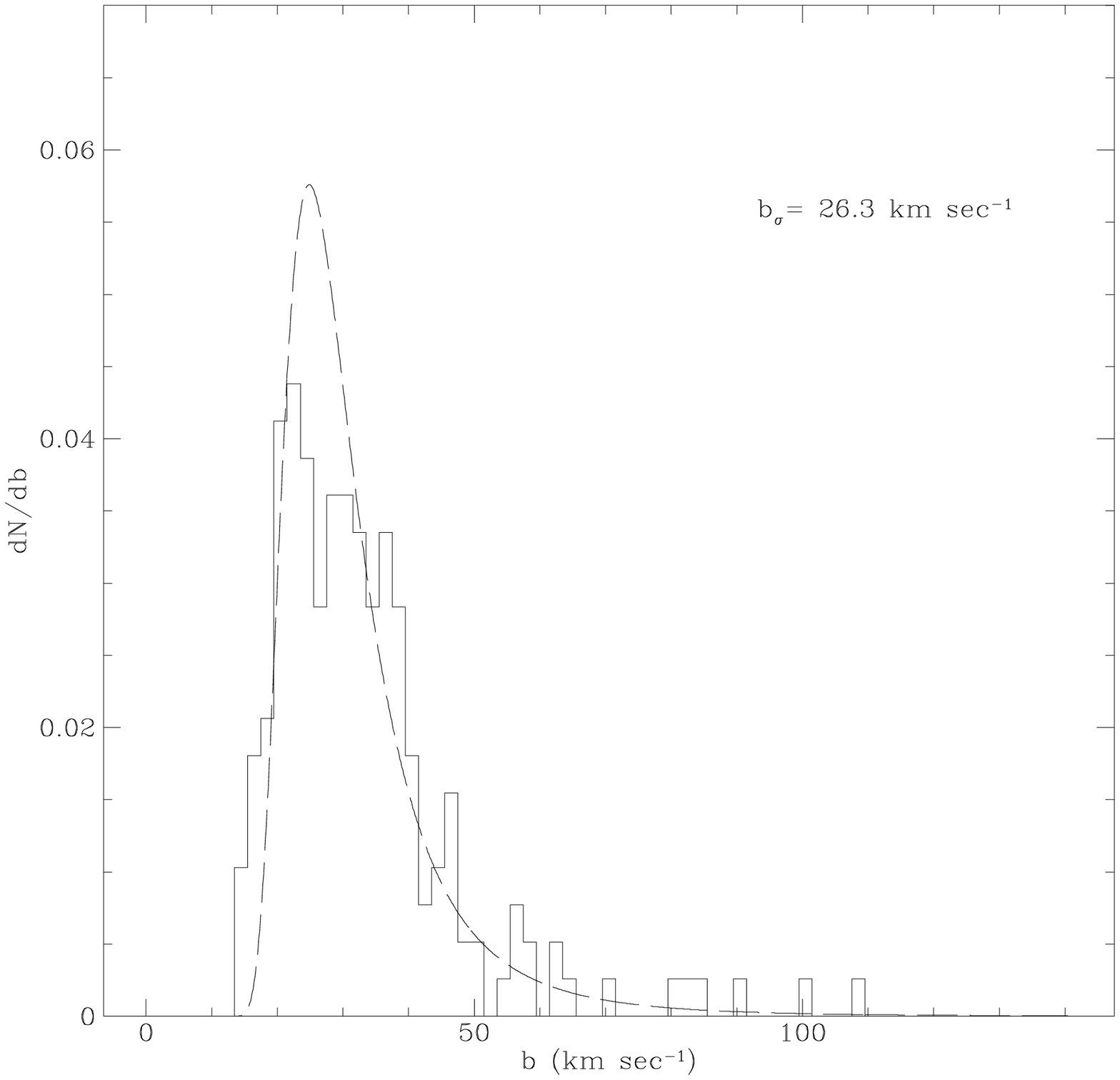 hoffset=-80 voffset=-80}{14.7cm}{21.5cm}
\FigNum{\ref{fig:dist}}
\end{figure}

\newpage
\begin{deluxetable}{lccccc}
\tablewidth{42pc}
\tablecaption{$b$ Data-sets \label{tab:dataset}
 }
\tablehead{
\colhead{QSO	} 	& \colhead{$\langle z_{\rm lines} \rangle$}	& \colhead{No. of $b$ values}  	& \colhead{$b_\sigma$ (km \persec)}   		&  \colhead{Max(${\rm p_{KS}}$)}	& \colhead{$b_\sigma$(Max(${\rm p_{KS}}$))}  \nl
\colhead{}		& \colhead{}	& \colhead{}	& \colhead{[${\rm p}_{\rm KS}> 1$\% range]}	&  \colhead{}   			&  (km \persec)}
\startdata
HS 1946+7658(A)	& 2.7	& 328			& --			& $4\times10^{-3} $ 						&  26.1  \nl
Q0014+813(B)		& 2.95	& 194			&[25.5, 27.2]		& 0.17								&  26.3 \nl
Q0302-003(B)		& 2.86	& 198			&[25.7, 28.8]		& 0.25								&  27.5 \nl
Q0636+680(B)		& 2.75	& 219			&[23.3, 24.8]		& 0.09								&  24.2 \nl
Q0956+122(B)		& 2.85	& 176			&[24.7, 27.2]		& 0.20								&  26.3 \nl 
Q0000-26(C)		& 3.7	& 287			& -- 			& $3\times10^{-3} $ 						&  23.7 \nl
\tablerefs{
(A) \citenp{kirkman97}: (B) \citenp{hu95}:  (C) \citenp{lu96} \nl
Only lines with $10^{12.5} \leq N_{\rm HI} \leq 10^{14} \, {\rm cm^{-2}}$ are
included in our samples.}
\enddata
\end{deluxetable}

\end{document}